\documentclass[12pt]{iopart}
\usepackage{graphicx}
\usepackage{dcolumn}
\usepackage{bm}
\usepackage{amssymb}
\usepackage{amssymb}

\def\>{\rangle}

\begin{document}

\title{Using a squeezed field to protect two-atom entanglement against spontaneous emissions}

\author{Jing Zhang$^1$, Re-Bing Wu$^1$, Chun-Wen Li$^1$, Tzyh-Jong Tarn$^{2}$}

\address{$^{1}$Department of Automation, Tsinghua University, Beijing 100084,
P. R. China\\
$^{2}$Department of Electrical and Systems Engineering, Washington
University, St. Louis, MO 63130, USA}
\ead{jing-zhang@mail.tsinghua.edu.cn}

\begin{abstract}
Tunable interaction between two atoms in a cavity is realized by
interacting the two atoms with an extra controllable single-mode
squeezed field. Such a controllable interaction can be further
used to control entanglement between the two atoms against
amplitude damping decoherence caused by spontaneous emissions. For
the independent amplitude damping decoherence channel,
entanglement will be lost completely without controls, while it
can be partially preserved by the proposed strategy. For the
collective amplitude damping decoherence channel, our strategy can
enhance the entanglement compared with the uncontrolled case when
the entanglement of the uncontrolled stationary state is not too
large.

\end{abstract}

\pacs{03.67.Lx,03.67.Mn,03.67.Pp}
\maketitle

\section{Introduction}\label{s1}
Quantum
entanglement~\cite{Nielsen,Einstein,Wootters,Ren,Vicente,Hassan}
is a fundamental property of multi-body quantum systems that shows
the non-local feature of quantum states. Quantum entanglement has
been commonly recognized to be an essential physical resource in
the implementation of high-speed quantum computation and
high-security quantum communication.

Many efforts have been made to create entanglement between
decoupled quantum systems. One natural way is to introduce a
simple intermediate device~\cite{Zheng,Oh,Li,Peskin,Sainz}, e.g.,
a single-mode field or an additional particle, whose coherent
interactions with the systems lead to their indirect interactions
with each other. The intermediate device can also be measured to
extract information about the quantum systems for quantum feedback
controls~\cite{WangJ,Carvalho1,Mancini} to manipulate the
entanglement dynamics. One may also utilize a dissipative
environment~\cite{Braun,Plenio1,Benatti,Nicolosi}, e.g., a
collective decoherence environment, to generate entanglement,
interacted with which the system irreversibly decays to a
stationary entangled state.

However, in most circumstances, quantum entanglement tends to be
destructed in environments~\cite{Yu1,Yu2,Yu3,Almeida1}. For
example, independent decoherence channels always lead to
disentanglement~\cite{Carvalho2} that is not recoverable by local
operations and classical communications.

Generally, non-local operations are required to effectively
protect entanglement. However, a non-local Hamiltonian generated
from the internal interaction between quantum systems, e.g., the
dipole-dipole interaction between two atoms via the vacuum, is
sometimes not a good choice, because disentanglement can also be
induced by decoherence under these interactions.

This paper introduces a single-mode squeezed field in a quantum
cavity to realize non-local controllable interactions between two
identical atoms in the weak coupling regime. By altering the
parameter amplification coefficient of the squeezed field, one can
continuously adjust the coupling strengths between atoms, which
can be further used to control the final entanglement between the
two atoms in presence of decoherence. It should be pointed out
that there is another interesting work on coupling the two atoms
via the squeezed vacuum~\cite{Ficek2}. Compared with the squeezed
vacuum, the auxiliary squeezed field in the cavity is more
controllable, which would be helpful to control the stationary
concurrence.

The paper is organized as follows: the physical model applied in
the paper is formulated in Sec.~\ref{s2}. Entanglement control
strategies are discussed for two-atom independent amplitude
damping decoherence channels, collective amplitude damping
decoherence channels, and their mixture, respectively in
Sec.~\ref{s3}, \ref{s4} and \ref{s5}. Conclusions and a forecast
of the future work are drawn in Sec.~\ref{s6}.

\section{Model Formulation}\label{s2}

Consider the system of two identical two-level atoms interacting
with a squeezed single-mode field in a quantum cavity (see
Fig.~\ref{Fig of two atoms interacting with a squeezed field in a
cavity}). \setlength{\mathindent}{0.5cm}
\begin{figure}[h]
\centerline{\includegraphics[width=2.8in,height=2.5in]{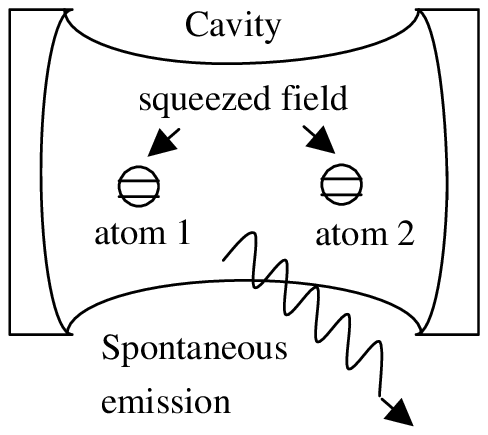}}\caption{Two
atoms undergoing decoherence caused by spontaneous emissions
interact with a single-mode squeezed field in a cavity.}\label{Fig
of two atoms interacting with a squeezed field in a cavity}
\end{figure}
The total Hamiltonian of the atoms and the cavity mode can be
described as below with $\hbar$ assumed to be $1$ without loss of
generality: \setlength{\mathindent}{3cm}
\begin{eqnarray}\label{Total Hamiltonian of the two atoms and the squeezed cavity mode}
H_{AC}&=&\omega_c
a^{\dagger}a+\frac{\omega_a}{2}\sum_{i=1}^2\sigma_z^{(i)}+\sum_{i=1}^2(\epsilon^{(i)}a\sigma_+^{(i)}+\epsilon^{(i)*}a^{\dagger}\sigma_-^{(i)})\nonumber\\
&&+(\xi e^{-i\Omega t} a^{\dagger\,2}+\xi^{*}e^{i\Omega t} a^2),
\end{eqnarray}
where the first two terms describe the free Hamiltonians of the
cavity mode and the atoms; $\omega_c$ is the frequency of the
cavity mode and $\omega_a$ is the inherent frequency of the atom
corresponding to the energy separation between the ground state
and the excited state of each atom; $a$ is the annihilation
operator of the cavity mode and $\sigma_z^{(i)},\,i=1,2$ is the
$z$-axis Pauli operator of the $i$-th atom. The third term
represents the interaction between the atoms and the cavity mode,
in which $\sigma_{\pm}^{(i)}=\sigma_x^{(i)}\pm
i\sigma_y^{(i)},\,i=1,2$, are the ladder operators of the $i$-th
atom. The complex coefficient
$$
\epsilon^{(i)}=\vec{\mu}\cdot \vec{g}(\vec{r}^{(i)}),
$$
is the inner product of the transition dipole moment $\vec{\mu}$
of each atom and the coupling constant
\setlength{\mathindent}{5.5cm}
\begin{equation}\label{Coupling constant between individual atom and the field}
\vec{g}(\vec{r}^{(i)})=\left(\frac{\omega_c}{2\epsilon_0
V}\right)^{\frac{1}{2}}\hat{e}_{\vec{k}}e^{i\vec{k}\cdot\vec{r}^{(i)}},
\end{equation}
where $\vec{r}^{(i)}$ is the position of the $i$-th atom;
$\vec{k}$ and $\hat{e}_{\vec{k}}$ are the wave vector and unit
polarization vector of the cavity mode; and $V$ is the
normalization volume of the cavity mode. The last term is the
Hamiltonian of the squeezed cavity mode, where the parameter
amplification coefficient $\xi$ and the frequency $\Omega$ are
continuously tunable. Such a manipulable standing squeezed field
in a high-Q cavity is realizable by squeezed state engineering
developed recently~\cite{Villas-Boas,Almeida2}. Roughly speaking,
a three-level atom in a ladder configuration is introduced to
interact with the cavity mode. In addition, a classical field is
used to manipulate the three-level atom, through which one can
continuously adjust the squeezed coefficient $\xi$ and the
frequency $\Omega$.

In the weak coupling regime, i.e., $\Delta=\omega_a-\omega_c$ and
$|\xi|\gg|\epsilon^{(i)}|$, $H_{AC}$ can be diagonalized by the
following unitary transform~\cite{Blais}:
$$
U=\exp\left[\frac{1}{\Delta}\sum_{i=1}^2(\epsilon^{(i)}a\sigma_+^{(i)}-\epsilon^{(i)*}a^{\dagger}\sigma_-^{(i)})\right],
$$
which, by taking the first-order approximation of
$\epsilon^{(i)}/\Delta$, gives the following expression:
\setlength{\mathindent}{1.5cm}
\begin{eqnarray*}
H_{AC}\approx UH_{AC}U^{\dagger}&\approx&\omega_c a^{\dagger}a+\xi
e^{-i\Omega
t}a^{\dagger\,2}+\xi^* e^{i\Omega t}a^2\\
&&+\sum_{i=1}^2\left[\frac{\tilde{\omega}_a}{2}+\frac{4|\epsilon^{(i)}|^2}{\Delta^2}\left(\xi
e^{-i\Omega t}a^{\dagger\,2}+{\rm
h.c.}\right)+\frac{4|\epsilon^{(i)}|^2}{\Delta}a^{\dagger}a\right]\sigma^{(i)}_z\\
&&+\sum_{i=1}^2\left[\left(\frac{2\epsilon^{(i)}\xi e^{-i\Omega
t}}{\Delta}a^{\dagger}+\frac{|\epsilon^{(i)}|^2\xi e^{-i\Omega
t}}{\Delta^2}\right)\sigma_+^{(i)}+{\rm h.c.}\right]\\
&&+\left(\mu_1 e^{-i(\Omega
t+\phi_1)}\sigma_+^{(1)}\sigma_+^{(2)}+{\rm h.c.}\right)\\
&&+\left(\mu_2 e^{-i\phi_2}\sigma_+^{(1)}\sigma_-^{(2)}+{\rm
h.c.}\right),
\end{eqnarray*}
\setlength{\mathindent}{1.5cm} where ${\rm h.c.}$ refers to
Hermitian conjugate;
$$
\tilde{\omega}_a=\omega_a+\frac{2\left(|\epsilon^{(1)}|^2+|\epsilon^{(2)}|^2\right)}{\Delta},\,\,\,\mu_1
e^{-i\phi_1}=\frac{2\xi\epsilon^{(1)}\epsilon^{(2)}}{\Delta^2},\,\,\,\mu_2
e^{-i\phi_2}=\frac{\epsilon^{(1)}\epsilon^{(2)*}}{\Delta}.$$

Further, by adiabatically eliminating the degrees of freedom of
the cavity mode, the following reduced two-atom Hamiltonian can be
obtained:
$$
H_A=\frac{\omega_a}{2}\sum_{i=1}^2\sigma_z^{(i)}+\left(\mu_1
e^{-i(\Omega t+\phi_1)}\sigma_+^{(1)}\sigma_+^{(2)}+{\rm
h.c.}\right)+\left(\mu_2
e^{-i\phi_2}\sigma_+^{(1)}\sigma_-^{(2)}+{\rm h.c.}\right),
$$
where the terms of individual atomic interaction with the cavity
are omitted due to the fact that
$$
\frac{\omega_a}{2}{\gg} |\epsilon^{(i)}|^2/\Delta,\,|\xi
\epsilon^{(i)}|/\Delta
$$
under the large detuning condition $\Delta\gg \epsilon^{(i)}$.
Since the parameter amplification coefficient $\xi$ and the
frequency $\Omega$ are tunable parameters, we have two control
parameters $\mu_1$ and $\Omega$ in $H_A$. In the interaction
picture, $H_A$ can be expressed as: \setlength{\mathindent}{1cm}
\begin{eqnarray}\label{Squeezed field induced two atom effective Hamiltonian}
H_A^{\rm eff}&=&\left(\mu_1
e^{-i\phi_1}\sigma_+^{(1)}\sigma_+^{(2)}+\mu_1
e^{i\phi_1}\sigma_-^{(1)}\sigma_-^{(2)}\right)+\left(\mu_2
e^{-i\phi_2}\sigma_+^{(1)}\sigma_-^{(2)}+\mu_2
e^{i\phi_2}\sigma_-^{(1)}\sigma_+^{(2)}\right),
\end{eqnarray}
\setlength{\mathindent}{3cm} when the parameter $\Omega$ is fixed
to be $2\omega_a$.

Besides the cavity mode, the atoms also interact with other modes
in the environment, which leads to the atomic spontaneous
emissions. In the case that the environmental modes are at the
vacuum state, the dynamics of atoms can be described by the
following master equation~\cite{Nicolosi,Ficek}:
\setlength{\mathindent}{2cm}
\begin{equation}\label{Master equation of atoms undergoing spontaneous emissions}
\dot{\rho}=-i[H_A^{\rm eff}+H_{12},\rho]+\sum_{i,j=1}^2
\Gamma_{ij}\left(\sigma_-^{(i)}\rho\sigma_+^{(j)}-\frac{1}{2}\rho\sigma_+^{(j)}\sigma_-^{(i)}-\frac{1}{2}\sigma_+^{(j)}\sigma_-^{(i)}\rho\right).
\end{equation}
The parameters \setlength{\mathindent}{5.5cm}
\begin{equation}\label{Inidividual spontaneous rates}
\Gamma_{11}=\Gamma_{22}=\Gamma=\frac{\omega_a^3\mu^2}{3\pi\epsilon_0
c^3}
\end{equation}
are the spontaneous emission rates of the individual atoms, where
$\mu=|\vec{\mu}|$ is the magnitude of the transition dipole
momentum, while
\begin{equation}\label{Collective spontaneous rates}
\Gamma_{12}=\Gamma_{21}=\Gamma F(k_0 r_{12})
\end{equation}
represent the collective spontaneous emission rates induced by the
coupling between the atoms. The function $F(k_0 r_{12})$ can be
expressed as~\cite{Nicolosi,Ficek}:
$$
F(k_0
r_{12})=\frac{3}{2}\left\{(1-3\cos^2\theta)\left[\frac{\cos(k_0
r_{12})}{(k_0 r_{12})^2}-\frac{\sin(k_0 r_{12})}{(k_0
r_{12})^3}\right]+\sin^2\theta\frac{\sin(k_0 r_{12})}{k_0
r_{12}}\right\},
$$
where $k_0=\omega_a/c$ and $\theta$ is the angle between the
dipole moment vector $\vec{\mu}$ and the vector
$\vec{r}_{12}=\vec{r}^{(1)}-\vec{r}^{(2)}$;
$r_{12}=|\vec{r}_{12}|$ is the distance between the two atoms. The
spontaneous emission process also introduces an additional
coherent dipole-dipole interaction between the atoms:
$$
H_{12}=\eta(\sigma_+^{(1)}\sigma_-^{(2)}+\sigma_-^{(1)}\sigma_+^{(2)}),
$$
where the coefficient $\eta$ in $H_{12}$ can be written
as~\cite{Ficek}: \setlength{\mathindent}{2cm}
\begin{eqnarray}\label{Interaction coefficient of the inherent dipole-dipole interaction between atoms}
\eta&=&\frac{3}{4}\Gamma\left\{[1-3\cos^2\theta]\left[\frac{\sin(k_0
r_{12})}{(k_0 r_{12})^2}+\frac{\cos(k_0 r_{12})}{(k_0
r_{12})^3}\right]-\sin^2\theta\frac{\cos(k_0 r_{12})}{k_0
r_{12}}\right\}.
\end{eqnarray}

\section{Independent amplitude damping decoherence
channel}\label{s3}

When the distance $r_{12}$ between the two atoms is far greater
than the resonant wavelength $1/k_0=c/\omega_a$ of the atom, i.e.,
$k_0 r_{12}\rightarrow\infty$, the amplitude damping decoherence
of the two atoms can be taken independently. Consequently, from
Eqs.~(\ref{Collective spontaneous rates}) and (\ref{Interaction
coefficient of the inherent dipole-dipole interaction between
atoms}), we have $\eta,\,\Gamma_{12},\,\Gamma_{21}\rightarrow 0$,
from which the following master equation holds:
\setlength{\mathindent}{4.5cm}
\begin{equation}\label{Master equation of independent amplitude damping decoherence channel}
\dot{\rho}=-i[H_A^{\rm eff},\rho]+\Gamma
\mathbb{D}[\sigma_-^{(1)}]\rho+\Gamma
\mathbb{D}[\sigma_-^{(2)}]\rho,
\end{equation}
where the superoperator $\mathbb{D}[L]\rho$ is defined as:
$$
\mathbb{D}[L]\rho=L\rho
L^{\dagger}-\frac{1}{2}L^{\dagger}L\rho-\frac{1}{2}\rho
L^{\dagger}L,
$$
and the two Lindblad terms $\mathbb{D}[\sigma_-^{(1)}]\rho$,
$\mathbb{D}[\sigma_-^{(2)}]\rho$ represent the amplitude damping
decoherence channels acting on the two atoms with the damping rate
$\Gamma>0$.

To measure the quantum entanglement, we use the
concurrence~\cite{Wootters} between the two atoms of the quantum
state $\rho$:
\begin{equation}\label{Concurrence}
C(\rho)=\max\{\lambda_1-\lambda_2-\lambda_3-\lambda_4,0\},
\end{equation}
where $\lambda_i's$ are the square roots of the eigenvalues, in
decreasing order, of the matrix:
$$
M=\rho(\sigma_y^{(1)}\sigma_y^{(2)})\rho^*(\sigma_y^{(1)}\sigma_y^{(2)}),
$$
and $\rho^*$ is the complex conjugate of $\rho$.

It is known that, in absence of the squeezed field, a two-atom
system will always be disentangled under independent amplitude
damping decoherence channels (see, e.g., Ref.~\cite{Carvalho2}),
and this is not recoverable by any local operations. However, the
entanglement can be partially protected via the intermediate
squeezed field, because the solution $\rho(t)$ of Eq.~(\ref{Master
equation of independent amplitude damping decoherence channel})
tends to a stationary state
\begin{equation}\label{Controlled stationary state under independent amplitude damping decoherence channels}
\rho_{\infty}=\frac{2\mu_1\Gamma}{4\mu_1^2+\Gamma^2}\rho_m+\left(1-\frac{2\mu_1\Gamma}{4\mu_1^2+\Gamma^2}\right)\rho_s,
\end{equation}
as a convex combination of a pure maximally entangled state
\begin{equation}\label{Maximally-entangled state induced by squeezed field}
\rho_m(\phi_1)=\frac{1}{2}\left(%
\begin{array}{cccc}
  1 &  &  & e^{-i\left(\phi_1-\frac{\pi}{2}\right)} \\
   & 0 &  &  \\
   &  & 0 &  \\
  e^{i\left(\phi_1-\frac{\pi}{2}\right)} &  &  & 1 \\
\end{array}%
\right)
\end{equation}
and a diagonal separable state
$$
\rho_s={\rm diag}(1-3\beta,\,\beta,\,\beta,\,\beta),
$$
where
$$
\beta=\frac{1}{8}\left(1-\sqrt{1-\left(\frac{4\mu_1\Gamma}{4\mu_1^2+\Gamma^2}\right)^2}\right).
$$
The subscript ``$m$" is an abbreviation of ``maximally entangled",
and the subscript ``$s$" refers to ``separable". The corresponding
stationary concurrence is:
\begin{eqnarray}\label{Stationary concurrence for independent amplitude damping decoherence channel}
C(\rho_{\infty})=\max\left\{\frac{2\mu_1(\Gamma-\mu_1)}{4\mu_1^2+\Gamma^2},0\right\},
\end{eqnarray}
which, when the coupling strength $\mu_1$ is tuned to be
$$
\mu_1=\frac{1}{\sqrt{5}+1}\Gamma,
$$
reaches its maximum value:
$$
C_{\rm max}=\frac{\sqrt{5}-1}{4}\approx 0.31>0.
$$
The plot of $C(\rho_{\infty})$ versus $\mu_1/\Gamma$ is shown in
Fig.~\ref{Fig of the stationary concurrence versus the control
parameters for independent amplitude damping decoherence}.
\setlength{\mathindent}{5cm}
\begin{figure}[h]
\centerline{\includegraphics[width=3.6in,height=2.2in]{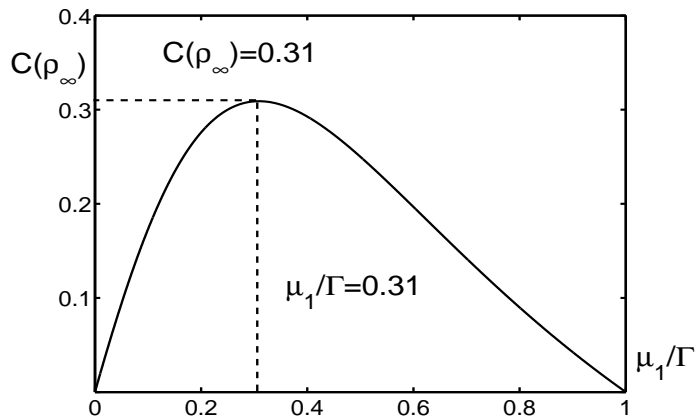}}\caption{Plot
of $C(\rho_{\infty})$ versus $\mu_1/\Gamma$.}\label{Fig of the
stationary concurrence versus the control parameters for
independent amplitude damping decoherence}
\end{figure}
\setlength{\mathindent}{4.5cm}

The cumbersome proof of Eqs. (\ref{Controlled stationary state
under independent amplitude damping decoherence channels}) and
(\ref{Stationary concurrence for independent amplitude damping
decoherence channel}) is shown in \ref{Proof of the results for
the independent amplitude damping decoherence}. We adopt here an
ideal model in which the two atoms have precise positions
$\vec{r}^{(i)}$. In real systems, position fluctuations are always
presented, i.e., \setlength{\mathindent}{6.5cm}
$$
\tilde{\vec{r}}^{(i)}=\vec{r}^{(i)}+\delta\vec{r}^{(i)},\,\,\,i=1,2,
$$
\setlength{\mathindent}{4.5cm}where $\tilde{\vec{r}}^{(i)}$ is the
actual position of the $i$-th atom and $\delta\vec{r}^{(i)}$ is the
corresponding fluctuation. From Eq.~(\ref{Coupling constant between
individual atom and the field}), the actual coupling coefficients
should be
$$
\tilde{\mu}_1
e^{i\tilde{\phi}_1}=e^{i\vec{k}\cdot(\delta\vec{r}^{(1)}+\delta\vec{r}^{(2)})}\mu_1
e^{i\phi_1}=\mu_1
e^{i(\phi_1+\vec{k}\cdot(\delta\vec{r}^{(1)}+\delta\vec{r}^{(2)}))},
$$which, consequently, fluctuates the phase by
\setlength{\mathindent}{4.5cm}
\begin{eqnarray}\label{Disturbed interaction strength and phase}
\delta\phi_1=\tilde{\phi}_1-\phi_1=\vec{k}\cdot(\delta\vec{r}^{(1)}+\delta\vec{r}^{(2)})=\delta\phi^{(1)}+\delta\phi^{(2)}
\end{eqnarray}
\setlength{\mathindent}{4.5cm}for the pure maximally entangled
state $\rho_m$ (which now it should be
$\rho_m(\tilde{\phi}_1)=\rho_m(\phi_1+\delta\phi_1)$) due to the
fluctuations of the positions of the atoms. Assume that
$\delta\phi^{(1)}$ and $\delta\phi^{(2)}$ obey Gaussian
distributions with means $0$ and variances $2\gamma_1$ and
$2\gamma_2$, one can verify by averaging over the random
fluctuations that the pure maximally entangled state $\rho_m$ is
blurred into a mixed state: \setlength{\mathindent}{2cm}
\begin{eqnarray*}
\bar{\rho}_m&=&\int_{-\infty}^{+\infty}\frac{d\delta\phi^{(1)}}{\sqrt{4\pi\gamma_1}}e^{-(\delta\phi^{(1)})^2/4\gamma_1}\int_{-\infty}^{+\infty}\frac{d\delta\phi^{(2)}}{\sqrt{4\pi\gamma_2}}e^{-(\delta\phi^{(2)})^2/4\gamma_2}\rho_m(\tilde{\phi}_1)\\
&=&\frac{1}{2}\left(
\begin{array}{cccc}
  1 &  &  & e^{-(\gamma_1+\gamma_2)}e^{-i\left(\phi_1-\frac{\pi}{2}\right)} \\
   & 0 &  &  \\
   &  & 0 &  \\
  e^{-(\gamma_1+\gamma_2)}e^{i\left(\phi_1-\frac{\pi}{2}\right)} &  &  & 1 \\
\end{array}
\right).
\end{eqnarray*}
\setlength{\mathindent}{4.5cm}Apparently, the resulting
entanglement is also reduced. In fact, in this case, the
stationary state should be
$$
\bar{\rho}_{\infty}=\frac{2\mu_1\Gamma}{4\mu_1^2+\Gamma^2}\bar{\rho}_m+\left(1-\frac{2\mu_1\Gamma}{4\mu_1^2+\Gamma^2}\right)\rho_s,
$$
with a modified stationary concurrence
$$
C(\bar{\rho}_{\infty})=\left\{e^{-(\gamma_1+\gamma_2)}\frac{2\mu_1\Gamma}{4\mu_1^2+\Gamma^2}-\frac{2\mu_1^2}{4\mu_1^2+\Gamma^2},\,0\right\}.
$$
The corresponding maximum stationary concurrence can be further
calculated as:
$$
\bar{C}_{\rm
max}=\frac{1}{4}\left(\sqrt{4e^{-(\gamma_1+\gamma_2)}+1}-1\right).
$$
Obviously, we have $\bar{C}_{\rm max}>0$, which means that our
strategy is still valid compared with the case without the
squeezed field.

However, the maximum stationary concurrence is reduced by the
dephasing effects caused by the fluctuations of the positions of
the atoms. In order to estimate the influence of the fluctuations
on the stationary entanglement, it can be estimated from
Eq.~(\ref{Disturbed interaction strength and phase}) that:
$$
\gamma_i=\frac{1}{2}{\rm
var}(\delta\phi^{(i)})\leq\frac{1}{2}|\vec{k}|^2 {\rm var}\left(
|\delta\vec{r}^{(i)}|\right)=2\pi^2\left(\frac{\delta
r^{(i)}}{\lambda}\right)^2,
$$
where $\lambda$ is the wavelength of the field in the cavity;
${\rm var}(\delta\phi^{(i)})$ is the variance of
$\delta\phi^{(i)}$ and $\delta {r}^{(i)}=\sqrt{{\rm
var}(|\delta\vec{r}^{(i)}|)}$ represents the magnitude of the
position fluctuation for the $i$-th atom. Therefore, if one is
capable of trapping the atom in the cavity such that
\setlength{\mathindent}{7cm}
\begin{equation}\label{Magnitude of delta_ri compared with the length of the field in the cavity}
\delta{r}^{(i)}\ll \lambda,
\end{equation}
the dephasing coefficients $\gamma_i$ can be neglected. This is
possible under the present atom trapping and cooling technique
since the wavelength $\lambda$ is of the order of $\mu$m (see,
e.g., Refs.~\cite{Quadt,Boozer,Pinkse,Maunz}). In this case, the
perturbed maximum stationary concurrence $\bar{C}_{\rm max}$ is
deviated slightly from the ideal maximum stationary concurrence
$C_{\rm max}$ (e.g., $\bar{C}_{\rm max}/C_{\rm max}\approx 90\%$
when $\delta{r}^{(i)}/\lambda=0.05$ as assumed in
Ref.~\cite{Quadt}). We can also see the influence from the
following example with parameters given in Ref.~\cite{Boozer}, in
which the mass $m$ of the atom (Cs atom), the oscillating
frequency $\omega$ of the external freedom of the atom (which is
different from $\omega_a$), the effective temperature $T_{\rm
eff}$ of the atom, and the wavelength $\lambda$ of the field in
the cavity are given as: \setlength{\mathindent}{4.5cm}
\begin{eqnarray*}
&m=133\times1.66\times10^{-27}{\rm kg}\approx2.2\times10^{-25}{\rm kg},&\\
&\omega=2\pi\times0.53\times10^6{\rm Hz}\approx3.3\times10^6{\rm Hz},&\\
&T_{\rm eff}=1.3\times10^{-4}{\rm
K},\,\,\,\lambda\approx0.9\times10^{-6}{\rm m}.&
\end{eqnarray*}
Here, we choose an effective temperature $T_{\rm eff}$ of the atom
that is ten times greater than the lowest cooling temperature
($13\mu{\rm K}$) given in Ref.~\cite{Boozer}, under which the
position of the atom can be taken as a classical parameter because
$$
\hbar\omega\ll\frac{1}{2}k_B T_{\rm eff},
$$
where $k_B$ is the Boltzmann constant. In this case, the position
fluctuation of the atom can be estimated from
$$
\frac{1}{2}m\omega^2(\delta{r}^{(i)})^2=\frac{1}{2}k_BT_{\rm eff},
$$
which leads to
$$
\delta{r}^{(i)}=\sqrt{\frac{k_B T_{\rm
eff}}{m\omega^2}}=\sqrt{\frac{1.38\times10^{-23}\times1.3\times10^{-4}}{2.2\times10^{-25}\times3.3^2\times10^{12}}}\approx2.7\times10^{-8}{\rm
m}.
$$
Thus, we have
$$
\frac{\delta{r}^{(i)}}{\lambda}\approx\frac{2.8\times10^{-8}}{0.9\times10^{-6}}=0.03,
$$
from which it can be calculated that $\bar{C}_{\rm max}/C_{\rm
max}\approx 97\%$. \setlength{\mathindent}{4.5cm}

Although the stationary concurrence may not be strong enough to be
directly applied in quantum information processing and would be
deteriorated by dephasing effects caused by noises such as the
position fluctuations of the atoms, it is still hopeful to be used
for entanglement protection. In fact, the fidelity between the
stationary state $\bar{\rho}_{\infty}$ and the maximally entangled
state $\rho_m$ can be calculated as:
$$
F(\bar{\rho}_{\infty})={\rm
tr}(\bar{\rho}_{\infty}\rho_m)=\frac{e^{-(\gamma_1+\gamma_2)}\mu_1\Gamma-\mu_1^2}{4\mu_1^2+\Gamma^2}+\frac{1}{2}.
$$
Optimally, it should be
\begin{equation}\label{Optimal fidelity}
\bar{F}_{\rm
max}=\frac{1}{8}\left(\sqrt{4e^{-(\gamma_1+\gamma_2)}+1}-1\right)+\frac{1}{2}>\frac{1}{2}.
\end{equation}
Since the maximum fidelity $\bar{F}_{\rm max}$ is always larger
than $0.5$, we can, in principle, increase the stationary
entanglement by introducing additional entanglement purification
process~\cite{Bennett,Dur}.

To illustrate our proposal, let us discuss their applications in
some typical circumstances. Firstly, consider the initial states
$\rho_0^1$ at the maximally entangled state
$$
\rho_0^1=\frac{1}{2}\left(%
\begin{array}{cccc}
  1 &  &  & 1 \\
   & 0 &  &  \\
   &  & 0 &  \\
  1 &  &  & 1 \\
\end{array}%
\right),
$$
and $\rho_0^2$ at the mixed entangled state
$$
\rho_0^2=\left(%
\begin{array}{cccc}
  0.85 &  &  & 0.1 \\
   & 0.03 &  &  \\
   &  & 0.07 &  \\
  0.1 &  &  & 0.05 \\
\end{array}%
\right).
$$
Moreover, let $\Gamma=1/\tau_0$, where $\tau_0$ is the relaxing
time constant. Simulation results are shown in Fig. \ref{Fig of
the concurrence for the independent amplitude damping decoherence
model}. \setlength{\mathindent}{0.5cm}
\begin{figure}[h]
\centerline{
\includegraphics[width=3.0in,height=2.2in]{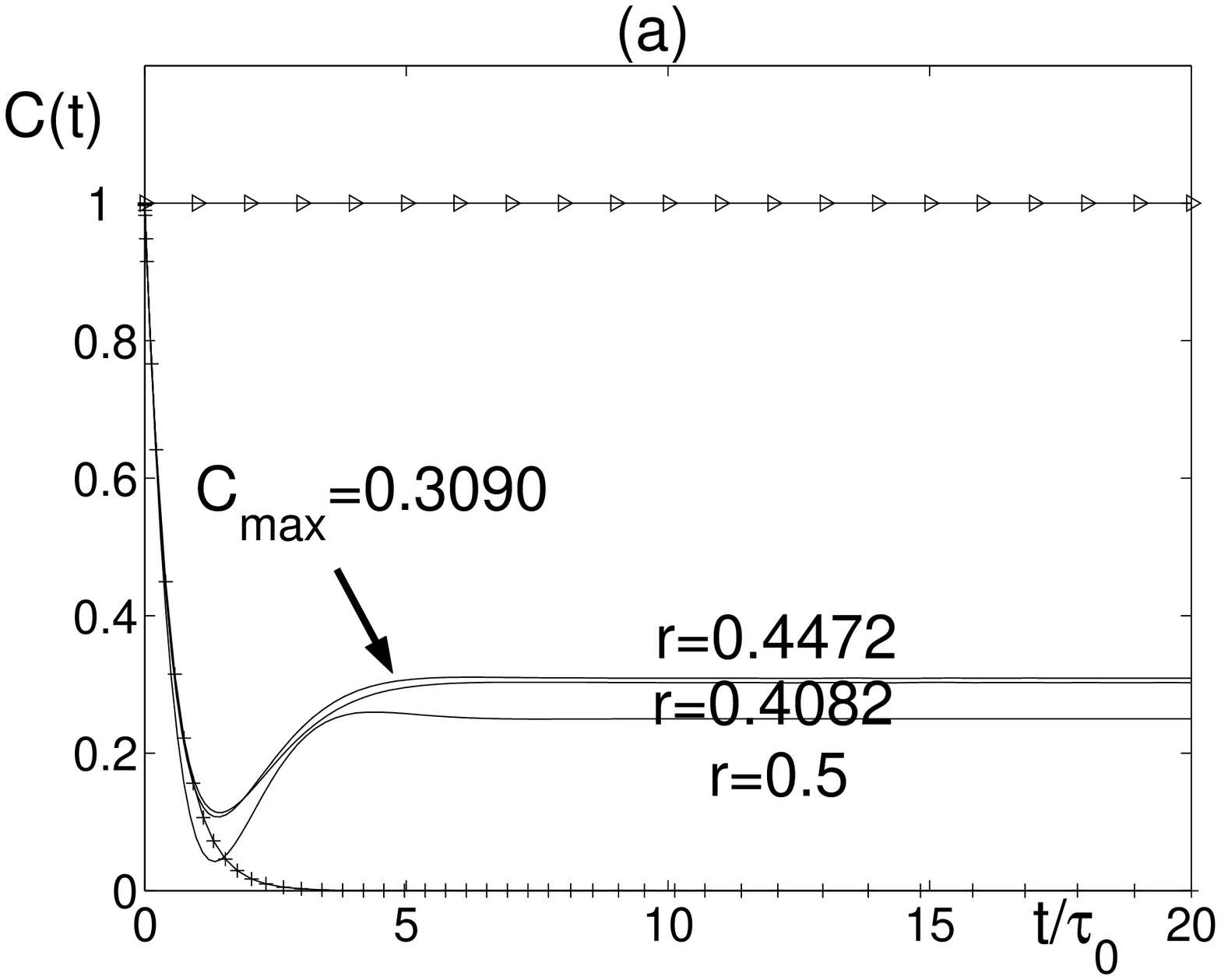}
\includegraphics[width=3.0in,height=2.2in]{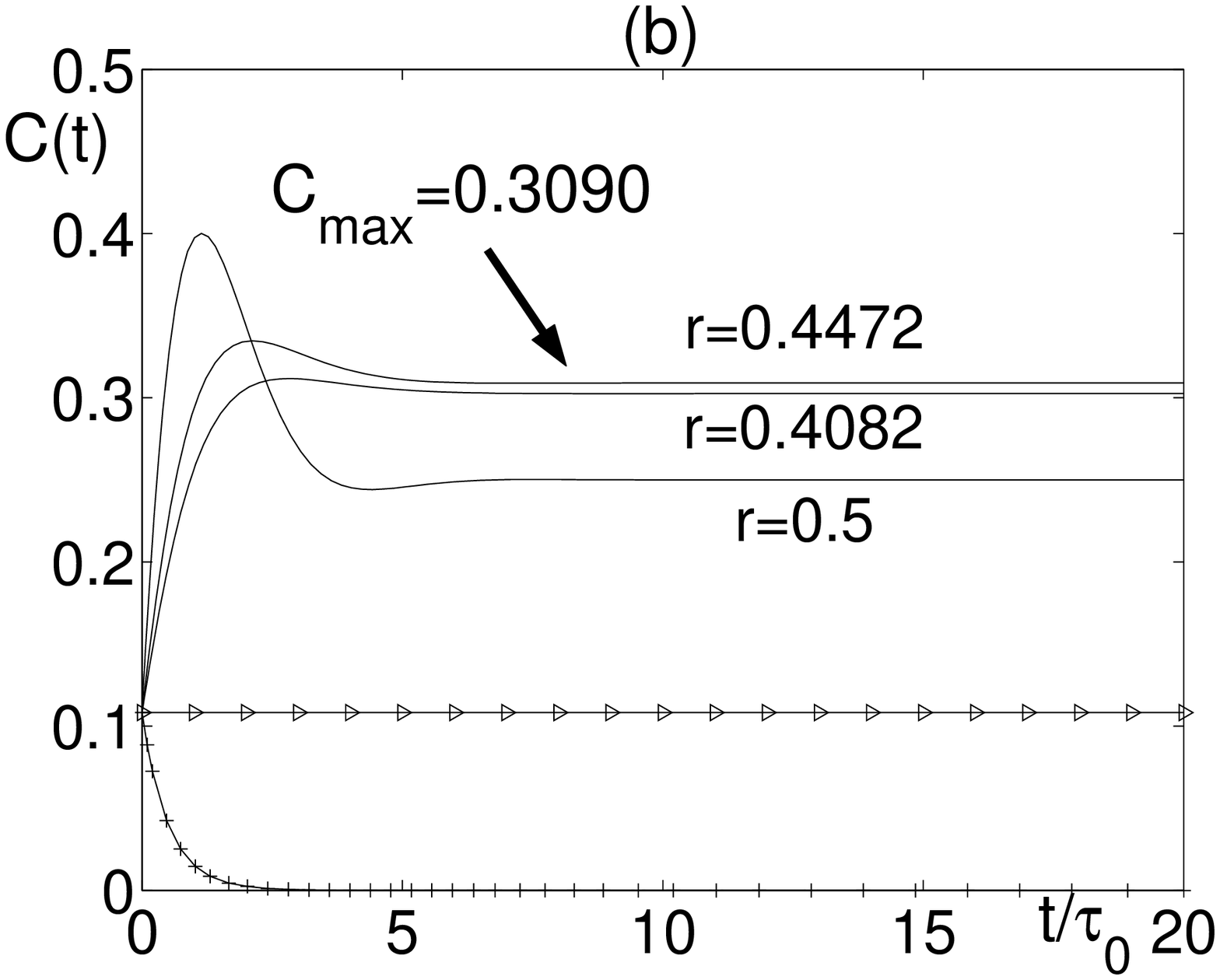}
} \caption{Plots of concurrence $C(t)$ where the initial states
are chosen as (a) maximally entangled state $\rho_0^1$ and (b)
mixed entangled state $\rho_0^2$. The plus-sign lines denote the
uncontrolled trajectories; the triangle lines are for the free
trajectories in absence of control and decoherence; the solid
lines represent the controlled trajectories with chosen parameters
$r=2\mu_1\Gamma/(4\mu_1^2+\Gamma^2)$, which is the proportion of
the maximally entangled state $\rho_m$ in Eq. (\ref{Controlled
stationary state under independent amplitude damping decoherence
channels}).}\label{Fig of the concurrence for the independent
amplitude damping decoherence model}
\end{figure}

It is shown in Fig. \ref{Fig of the concurrence for the
independent amplitude damping decoherence model} that the
entanglement of the quantum states always decays to zero without
control as what is known in the literature~\cite{Carvalho2}. The
corresponding stationary state is the two-atom ground state:
$$
\rho^u_{\infty}=|00\rangle\langle00|=\left(%
\begin{array}{cccc}
  1 &  &  &  \\
   & 0 &  &  \\
   &  & 0 &  \\
   &  &  & 0 \\
\end{array}%
\right),
$$
which is at the boundary of the set of all separable states. The
superscript ``$u$" refers to the ``uncontrolled" system. When our
strategy is applied, the entanglement can be remarkably retrieved
against decoherence, and the maximum concurrence of the stationary
state is
$$
C_{\max}=\frac{\sqrt{5}-1}{4}\approx 0.31,
$$
when
$$
r=\frac{2\mu_1\Gamma}{4\mu_1^2+\Gamma^2}=\frac{1}{\sqrt{5}}\approx
0.45.
$$

It is also noted that our strategy can enhance the entanglement of
the stationary state of the naked atoms (i.e., neither control nor
decoherence exist).

\section{Collective amplitude damping decoherence
channel}\label{s4}

When the distance between the atoms is far shorter than the
resonant wavelength of the atom, i.e., $k_0 r_{12}\rightarrow 0$,
from Eqs.~(\ref{Collective spontaneous rates}) and
(\ref{Interaction coefficient of the inherent dipole-dipole
interaction between atoms}) we have:
$$
\eta\rightarrow\eta_0=\frac{3\Gamma}{4k_0^3r_{12}^3}(1-3\cos^2\theta),\,\,\,\Gamma_{12}=\Gamma_{21}\rightarrow\Gamma,
$$
which corresponds to a two-atom collective amplitude damping
decoherence channel~\cite{Duan}. In this case, the master equation
of the two atoms becomes: \setlength{\mathindent}{3cm}
\begin{eqnarray}\label{Master equation of collective amplitude damping decoherence channel}
\dot{\rho}&=&-i\left[H_A^{\rm
eff}+\eta_0\left(\sigma_+^{(1)}\sigma_-^{(2)}+\sigma_-^{(1)}\sigma_+^{(2)}\right),\rho\right]+\Gamma\mathbb{D}[S_-]\rho,
\end{eqnarray}
\setlength{\mathindent}{5.5cm} where the two-atom operator
$S_-=\sigma_-^{(1)}+\sigma_-^{(2)}$, and $\Gamma>0$ is the damping
rate. Because the two atoms are very close to each other, from Eq.
(\ref{Coupling constant between individual atom and the field})
the coupling strength between each atom and the cavity can be
taken as identical, i.e.,
$\epsilon^{(1)}=\epsilon^{(2)}=\epsilon$, so that the interaction
Hamiltonian $H_A^{\rm eff}$ can be expressed as:
$$
H_A^{\rm eff}=\mu_1\left(
e^{-i\phi_1}\sigma_+^{(1)}\sigma_+^{(2)}+
e^{i\phi_1}\sigma_-^{(1)}\sigma_-^{(2)}\right)+\mu_2\left(\sigma_+^{(1)}\sigma_-^{(2)}+\sigma_-^{(1)}\sigma_+^{(2)}\right),
$$
where
$$
\mu_1
e^{-i\phi_1}=2\xi\epsilon^2/\Delta^2,\,\,\mu_2=|\epsilon|^2/\Delta.
$$

In absence of the intermediate squeezed field, i.e.,
$\mu_1=\mu_2=0$, the stationary state of the two-atom system
\begin{equation}\label{Stationary state of the uncontrolled collective amplitude damping decoherence channel}
\rho^u_{\infty}=(1-\kappa)\tilde{\rho}_m+\kappa\rho_0
\end{equation}
is a convex combination of the maximally entangled state
\begin{equation}\label{Maximally-entangled state induced by collective decoherence channel}
\tilde{\rho}_m=\frac{1}{2}\left(%
\begin{array}{cccc}
  0 &  &  &  \\
   & 1 & -1 &  \\
   & -1 & 1 &  \\
   &  &  & 0 \\
\end{array}%
\right),
\end{equation}
and the two-atom ground state
$$
\rho_0=\left(%
\begin{array}{cccc}
  1 &  &  &  \\
   & 0 &  &  \\
   &  & 0 &  \\
   &  &  & 0 \\
\end{array}%
\right),
$$
where the weight $\kappa\in[0,1]$ is determined by the initial
density matrix:
$$
\kappa={\rm
tr}\left[\left(\frac{1}{4}\sigma_z^{(1)}\sigma_z^{(2)}\right)\rho(t_0)\right]+\frac{\sqrt{2}}{2}{\rm
tr}(\Omega_{23}^x\rho(t_0))+\frac{3}{4},
$$
and
$$
\Omega_{23}^x=\frac{1}{\sqrt{2}}\left(%
\begin{array}{cccc}
   &  &  & 0 \\
   &  & 1 &  \\
   & 1 &  &  \\
  0 &  &  &  \\
\end{array}%
\right).
$$
The resulting stationary concurrence is
$$C(\rho^u_{\infty})=1-\kappa.$$

When the intermediate squeezed field is presented, the
corresponding two-atom stationary state
\setlength{\mathindent}{4.5cm}
\begin{equation}\label{Stationary state of the controlled collective amplitude damping decoherence channel}
\rho_{\infty}=s\tilde{\rho}_m+r\rho_m+(1-s-r)\tilde{\rho}_s
\end{equation}
is a convex combination of the maximally entangled states
$\tilde{\rho}_m$ and $\rho_m$ given in
Eqs.~(\ref{Maximally-entangled state induced by collective
decoherence channel}) and (\ref{Maximally-entangled state induced
by squeezed field}) respectively, and a diagonal separable state
$$
\tilde{\rho_s}={\rm
diag}\left(\tilde{\beta}_1+\tilde{\beta}_2,\frac{1}{2}-\tilde{\beta}_2,\frac{1}{2}-\tilde{\beta}_2,\tilde{\beta}_2-\tilde{\beta}_1\right),
$$
where
$$
\tilde{\beta}_1=\frac{\kappa\Gamma^2}{2(\Gamma^2+3\mu_1^2)},\quad\tilde{\beta}_2=\frac{\kappa(\Gamma^2+2\mu_1^2)}{2(\Gamma^2+3\mu_1^2)}.
$$
The weights $s$ and $r$ are, respectively,
$$s=1-\frac{7}{6}\kappa+\frac{\Gamma^2-3\mu_1^2}{\Gamma^2+3\mu_1^2}\kappa,\,\,\,r=\frac{2\Gamma\mu_1\kappa}{\Gamma^2+3\mu_1^2}.$$
It can be examined that when the parameter $\mu_1$ is in the
range: \setlength{\mathindent}{3.5cm}
\begin{equation}\label{Interval of the control parameter}
\frac{\kappa-\sqrt{-9\kappa^2+22\kappa-12}}{6-5\kappa}\leq\frac{\mu_1}{\Gamma}\leq\frac{\kappa+\sqrt{-9\kappa^2+22\kappa-12}}{6-5\kappa},
\end{equation}
the resulting stationary concurrence $C(\rho_{\infty})$ is
superior to $C(\rho^u_{\infty})$ without the intermediate squeezed
field:
\begin{equation}\label{Controlled stationary concurrence under collective amplitude damping decoherence}
C(\rho_{\infty})=\frac{(\Gamma^2+2\mu_1^2+2\Gamma\mu_1)\kappa}{\Gamma^2+3\mu_1^2}-1\geq
C(\rho^u_{\infty})=1-\kappa.
\end{equation}

The interval given in (\ref{Interval of the control parameter}) is
nonempty if and only if \setlength{\mathindent}{6cm}
\begin{equation}\label{Interval of the dissipation parameter}
\frac{11}{9}-\frac{1}{9}\sqrt{13}\leq\kappa\leq 1,
\end{equation}
otherwise, our strategy is not capable of improving the stationary
concurrence. Moreover, the maxima of $C(\rho_{\infty})$ is
achieved when
$$
\mu_1=\frac{2}{\sqrt{13}+1}\Gamma,
$$
and the corresponding maximum value is
$$
C_{\max}=\frac{\sqrt{13}+5}{6}\kappa-1.
$$
The plots of the stationary concurrence versus the coupling
strength for different $\kappa$ are shown in Fig.~\ref{Fig of the
stationary concurrence versus the control parameters for
collective amplitude damping decoherence}. In Fig.~\ref{Fig of the
stationary concurrence versus the control parameters for
collective amplitude damping decoherence}a, the controlled
stationary concurrence is superior to the uncontrolled one only in
the interval given by Eq.~(\ref{Interval of the control
parameter}), while, in Fig.~\ref{Fig of the stationary concurrence
versus the control parameters for collective amplitude damping
decoherence}b, the controlled stationary concurrence is always
better than the uncontrolled one.

\setlength{\mathindent}{0.5cm}
\begin{figure}[h]
\centerline{
\includegraphics[width=2.8in,height=2.2in]{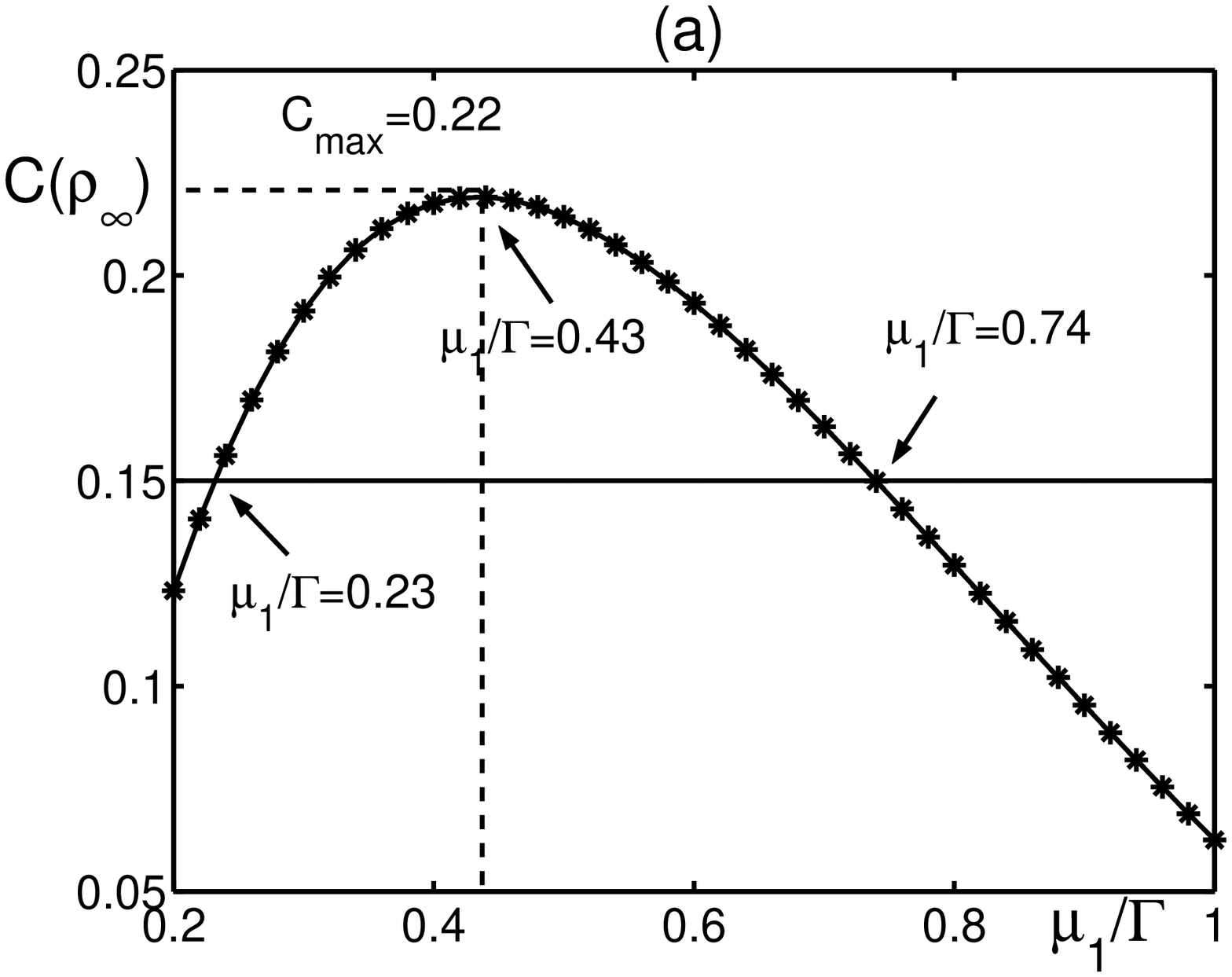}
\includegraphics[width=2.8in,height=2.2in]{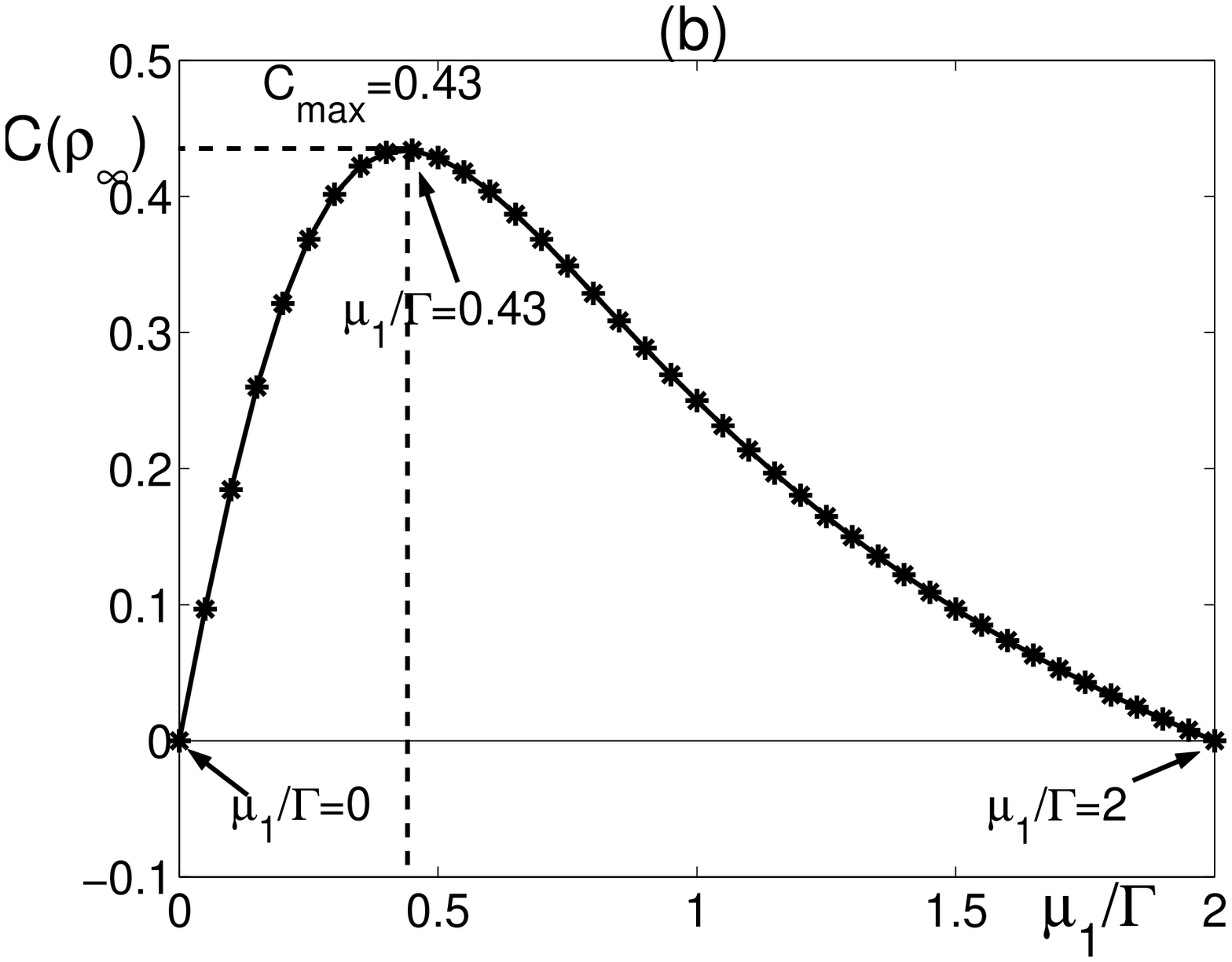}
} \caption{Plots of $C(\rho_{\infty})$ versus $\mu_1/\Gamma$ for
(a) $\kappa=0.85$, (b) $\kappa=1$. The asterisk line is for the
controlled stationary concurrence and the solid line is for the
uncontrolled stationary concurrence.}\label{Fig of the stationary
concurrence versus the control parameters for collective amplitude
damping decoherence}
\end{figure}

The fact that our strategy is effective only when the parameter
$\kappa$ is sufficiently large comes from the competition between
$\tilde{\rho}_m$ and $\rho_m$ in Eq.~(\ref{Stationary state of the
controlled collective amplitude damping decoherence channel}),
where $\tilde{\rho}_m$ comes from the dissipation effect and
$\rho_m$ is induced by our proposal. When $\kappa$ is close to
$0$, the dissipation dominates and hence the control fails, while,
when $\kappa$ is close to $1$, the control becomes effective.

As has been indicated in Sec.~\ref{s3}, the fluctuations of the
positions of the atoms would bring an uncertain phase shift for
the maximally entangled state $\rho_m$, which may deteriorate the
stationary concurrence. The calculations are like those in
Sec.~\ref{s3}, so we omit them here.

Fig. \ref{Fig of the concurrence for the collective amplitude
damping decoherence model} shows some numerical examples, where
the initial states are, respectively,
$$
\rho_0^1=\left(%
\begin{array}{cccc}
  3/8 &  &  & 3/8 \\
   & 1/8 &  &  \\
   &  & 1/8 &  \\
  3/8 &  &  & 3/8 \\
\end{array}%
\right),
$$
and
$$
\rho_0^2=\left(%
\begin{array}{cccc}
  0.85 &  &  & 0.1 \\
   & 0.03 &  &  \\
   &  & 0.07 &  \\
  0.1 &  &  & 0.05 \\
\end{array}%
\right).
$$
\setlength{\mathindent}{0.5cm}
\begin{figure}[h]
\centerline{
\includegraphics[width=3.0in,height=2.2in]{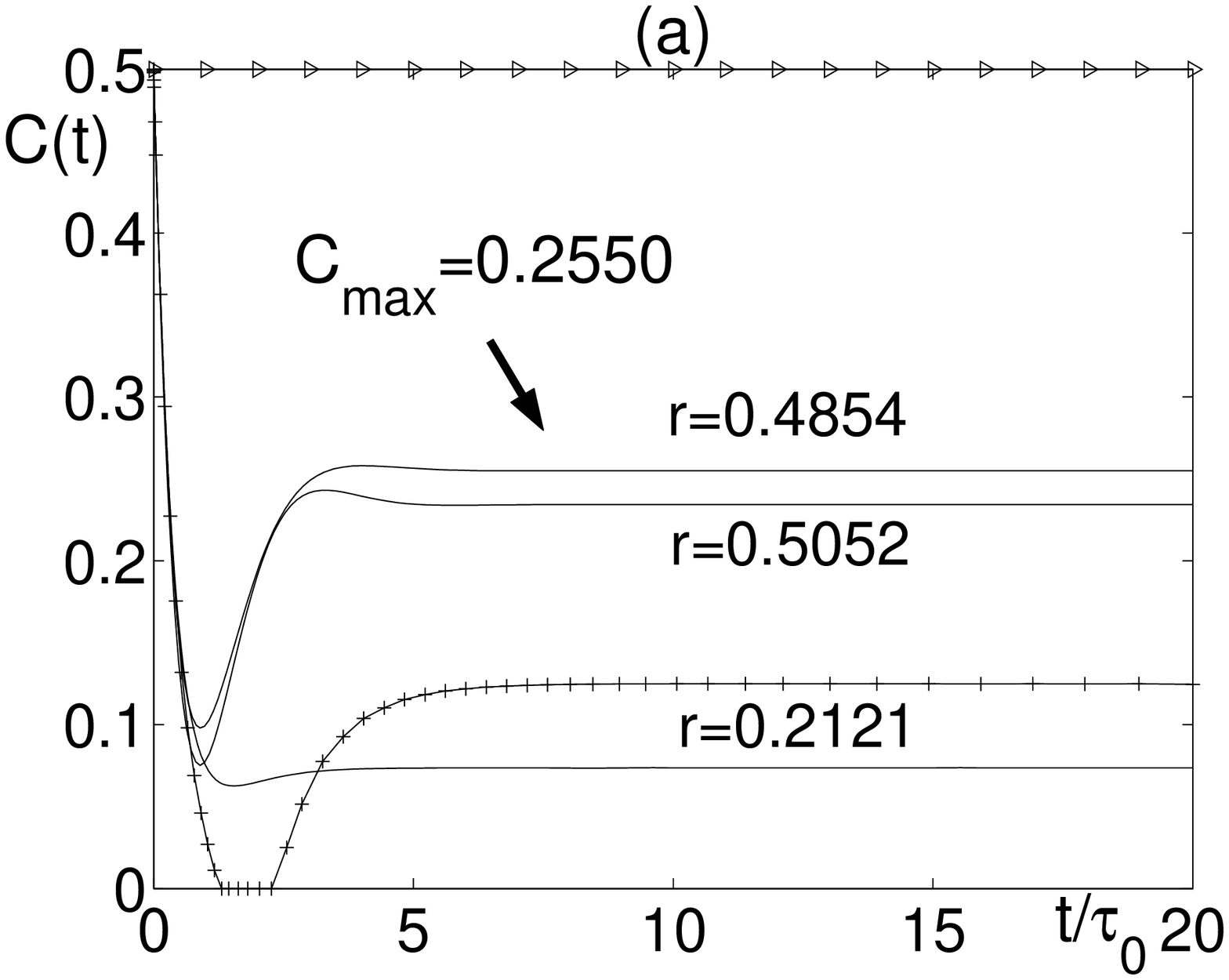}
\includegraphics[width=3.0in,height=2.2in]{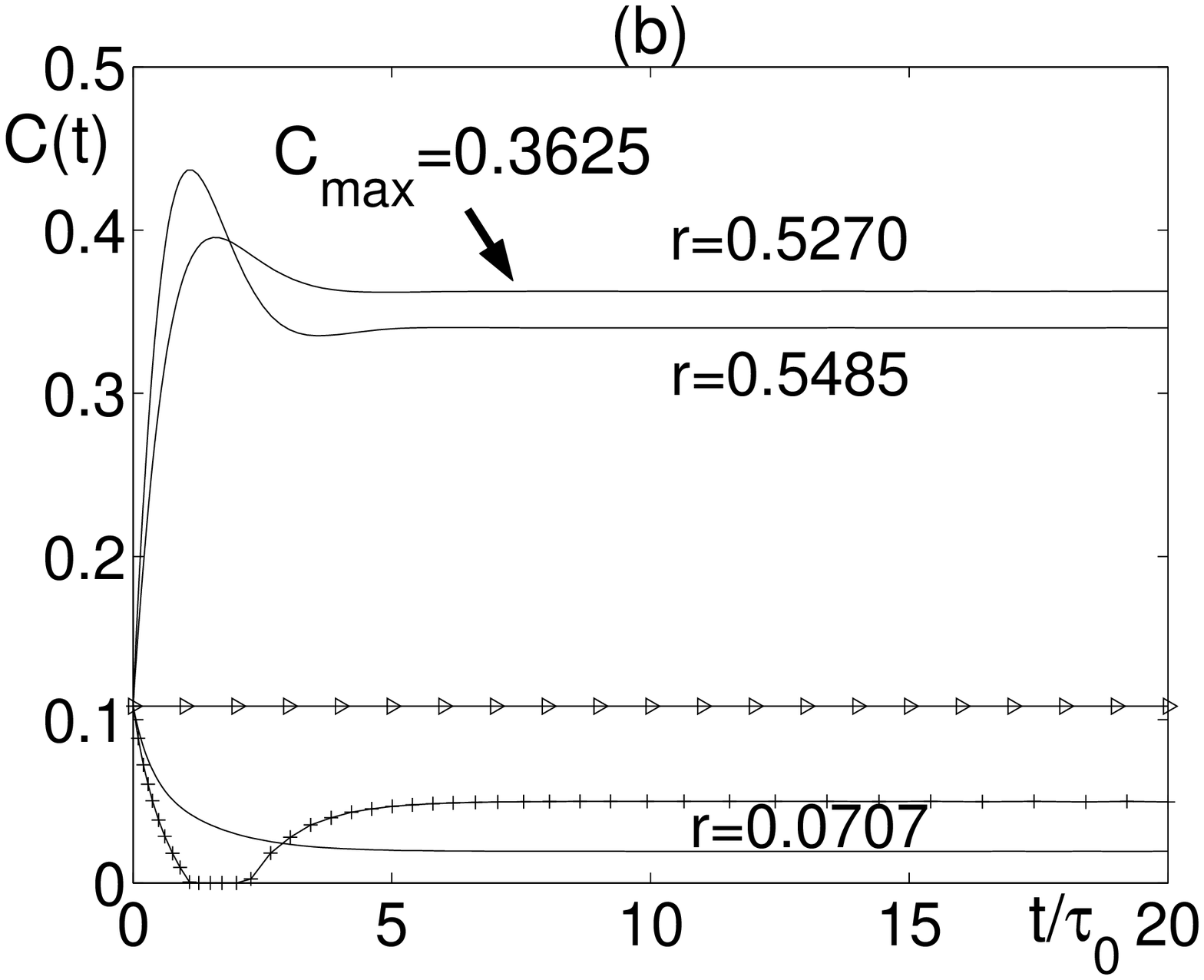}
} \caption{Plots of the concurrence $C(t)$ where the system state
is initialized for (a) $\rho_0^1$ and (b) $\rho_0^2$. The
plus-sign lines denote the uncontrolled trajectories; the triangle
lines are for the free trajectories in absence of control and
decoherence; the solid lines represent the controlled trajectories
with different parameters
$r=2\Gamma\mu_1\kappa/(\Gamma^2+3\mu_1^2)$, which is the
proportion of the squeezed field induced maximally entangled state
$\rho_m$ in Eq. (\ref{Stationary state of the controlled
collective amplitude damping decoherence channel}).}\label{Fig of
the concurrence for the collective amplitude damping decoherence
model}
\end{figure}

The simulation results show that the stationary state of the
uncontrolled system under the collective amplitude damping
decoherence may remain entangled which is quite different from the
independent decoherence channel, and this feature has been
utilized in the literature~\cite{Braun,Plenio1,Benatti} to create
entanglement between qubits. Our strategy may further increase the
entanglement in the stationary entangled state, as shown in
comparison between the plus-sign lines and the solid lines.

Another feature of the collective amplitude damping decoherence
channel observed from Fig. \ref{Fig of the concurrence for the
collective amplitude damping decoherence model} is that the
maximum concurrence depends on the initial state. For certain
values of $r$, our control strategy may have worse performance
than that induced by the natural dissipation. Both Fig.~\ref{Fig
of the concurrence for the collective amplitude damping
decoherence model}a and \ref{Fig of the concurrence for the
collective amplitude damping decoherence model}b provide such a
case where the solid line (controlled trajectory) goes below the
plus-sign line (uncontrolled trajectory). The corresponding
control parameter $\mu_1$ is outside the interval given in
Eq.~(\ref{Interval of the control parameter}).

\section{Mixed amplitude damping decoherence channel}\label{s5}

In actual experiments, the decoherence channel is never perfectly
collective, because it is hard to place the two atoms in an cavity
close enough. The existing atom trapping and cooling
techniques~\cite{Eichmann,DeVoe} can only hold two atoms
approximately at the distance of the same order of the resonant
wavelength of the atom. Thus, it is more realistic to treat the
resulting decoherence channel as a mixture of an independent
amplitude damping decoherence channel and a collective amplitude
damping decoherence channel, as shown in the following master
equation: \setlength{\mathindent}{2cm}
\begin{eqnarray}\label{Master equation for the general amplitude damping decoherence channel}
\dot{\rho}&=&-i[H_A^{\rm eff},\rho]+\sum_{i=1}^2\Gamma\mathcal{D}[\sigma_{-}^{(i)}]\rho+\Gamma_{12}\left(\sigma_-^{(1)}\rho\sigma_{+}^{(2)}-\frac{1}{2}\{\sigma_{+}^{(2)}\sigma_{-}^{(1)},\rho\}\right)\nonumber\\
&&+\Gamma_{12}\left(\sigma_{-}^{(2)}\rho\sigma_{+}^{(1)}-\frac{1}{2}\{\sigma_{+}^{(1)}\sigma_{-}^{(2)},\rho\}\right),
\end{eqnarray}
where $0<\Gamma_{12}<\Gamma$.

It can be verified that the stationary state of the uncontrolled
system is nothing but the separable two-atom ground state
$\rho^u_{\infty}=|00\rangle\langle 00|$ in which entanglement
completely disappears, as well as in the case of the independent
decoherence channel. By introducing the intermediate squeezed
field, we can stabilize the system at the same stationary state
given in Eq.~(\ref{Controlled stationary state under independent
amplitude damping decoherence channels}).

\section{Conclusions}\label{s6}

In summary, we proposed a two-atom entanglement control strategy,
via a controllable squeezed field coupled to the two atoms, to
protect entanglement from the spontaneous emission process. The
parameter amplification coefficient of the squeezed field can be
tuned to generate a non-local Hamiltonian, which can be used to
maintain entanglement of the two-atom states against decoherence.
For the independent amplitude damping decoherence channel, we can
partially recover the entanglement of the quantum state which
otherwise will be completely lost. For the collective amplitude
damping decoherence channel, our strategy can effectively enhance
the entanglement of the stationary state compared with the
dissipation-induced strategies provided that the uncontrolled
stationary state is not tightly entangled.

The proposed entanglement control strategy is an open-loop control
strategy, where no measurements are done during the course of
control. Such control strategies require exact values of the
system parameters, and can badly suffer from the uncertainty of
these parameters, which may bring remarkable derivation of the
stationary concurrence from the ideal values. This problem is
hopefully solvable by quantum feedback controls.

Another direction of the succeeding research will be the
application in solid state systems. In such systems, controllable
coupling between qubits is easier to be achieved~\cite{Liu1,Liu2}
compared with the optical systems. However, interactions between
the solid-state systems and their environments are more
complicated, which may lead to non-Markovian noises~\cite{Yu4}. To
what extent the controllable non-local unitary operations can
preserve entanglement against non-Markovian noises is an
interesting problem to be explored, for which existing decoherence
suppression strategies~\cite{Ganesan,MZhang,WCui} may be helpful.
\\[0.2cm]

\appendix
\section{Proof of the Equations (\ref{Controlled stationary state under independent amplitude damping decoherence channels}) and (\ref{Stationary concurrence for independent amplitude damping
decoherence channel})}\label{Proof of the results for the
independent amplitude damping decoherence}

Firstly, we transform the control model (\ref{Master equation of
independent amplitude damping decoherence channel}) from the
complex matrix space into the real vector space, i.e., the
so-called coherence vector
picture~\cite{Alicki,Altafini,Zhang3,Zhang1}. With respect to the
inner product $\langle X, Y\rangle={\rm tr}(X^{\dagger}Y)$, we
define the following orthonormal basis for all two-atom operators:
\setlength{\mathindent}{2.5cm}
\begin{eqnarray}\label{Matrix basis for two-qubit operators}
\left\{\frac{1}{2}I_{4 \times
4},\,\Omega_{14}^x,\,\Omega_{14}^y,\,\Omega_{23}^x,\,\Omega_{23}^y,\,\frac{1}{2}\sigma_x^{(1)},\,\frac{1}{2}\sigma_y^{(1)},\,\frac{1}{2}\sigma_x^{(2)},\frac{1}{2}\sigma_y^{(2)},\,\frac{1}{2}\sigma_x^{(1)}\sigma_z^{(2)},\right.\nonumber\\
\left.\frac{1}{2}\sigma_z^{(1)}\sigma_x^{(2)},\,\frac{1}{2}\sigma_y^{(1)}\sigma_z^{(2)},\,\frac{1}{2}\sigma_z^{(1)}\sigma_y^{(2)},\,\Omega_{14}^z,\Omega_{23}^z,\frac{1}{2}\sigma_z^{(1)}\sigma_z^{(2)}\right\},
\end{eqnarray}
where
$$
\Omega_{14}^x=\frac{1}{\sqrt{2}}\left(%
\begin{array}{cccc}
   &  &  & 1 \\
   &  & 0 &  \\
   & 0 &  &  \\
  1 &  &  &  \\
\end{array}%
\right),\,\,\Omega_{14}^y=\frac{1}{\sqrt{2}}\left(%
\begin{array}{cccc}
   &  &  & -i \\
   &  & 0 &  \\
   & 0 &  &  \\
  i &  &  & 0 \\
\end{array}%
\right),$$
$$
\Omega_{23}^x=\frac{1}{\sqrt{2}}\left(%
\begin{array}{cccc}
   &  &  & 0 \\
   &  & 1 &  \\
   & 1 &  &  \\
  0 &  &  &  \\
\end{array}%
\right),\,\,\Omega_{23}^y=\frac{1}{\sqrt{2}}\left(%
\begin{array}{cccc}
   &  &  & 0 \\
   &  & -i &  \\
   & i &  &  \\
  0 &  &  &  \\
\end{array}%
\right),$$
$$
\Omega_{14}^z=\frac{1}{\sqrt{2}}\left(%
\begin{array}{cccc}
  1 &  &  &  \\
   & 0 &  &  \\
   &  & 0 &  \\
   &  &  & -1 \\
\end{array}%
\right),\,\,\Omega_{23}^z=\frac{1}{\sqrt{2}}\left(%
\begin{array}{cccc}
  0 &  &  &  \\
   & 1 &  &  \\
   &  & -1 &  \\
   &  &  & 0 \\
\end{array}%
\right).
$$
Under this basis, the system density matrix can be expressed as:
$$
\rho=\frac{1}{4}I_{4 \times 4}+\sum_{i=1}^{15} m_i\Omega_i,
$$
where $m_i={\rm tr}(\Omega_i\rho)$ and $\Omega_i,\,i=1,\cdots,15$,
are the basis matrices in Eq.~(\ref{Matrix basis for two-qubit
operators}) except $\frac{1}{2} I_{4 \times 4}$.
$m=(m_1,\cdots,m_{15})^T$ is called the coherence vector of
$\rho$.

In the coherence vector picture, the master equation (\ref{Master
equation of independent amplitude damping decoherence channel})
can be rewritten as~\cite{Alicki,Altafini,Zhang3,Zhang1}:
\setlength{\mathindent}{6cm}
\begin{equation}\label{Coherence vector equation of independent amplitude damping decoherence channel}
\dot{m}=O_A m+Dm+g,
\end{equation}
where the orthogonal matrix $O_A$ is the adjoint
representation~\cite{Altafini} of $-iH_A^{\rm eff}$. The affine
term ``$Dm+g$" is that of the Lindblad terms:
$$
\Gamma\mathbb{D}[\sigma_-^{(1)}]\rho+\Gamma\mathbb{D}[\sigma_-^{(2)}]\rho,
$$
where $D\leq 0$ and $g$ is a constant vector. Further, divide $m$
into the following sub-vectors: \setlength{\mathindent}{4.5cm}
\begin{eqnarray}\label{mp meta and mepsilon}
m^p&=&(m_{14}^x,m_{14}^y,m_{23}^x,m_{23}^y)^T,\nonumber\\
m^{\eta}&=&(m_{14}^z,m_{23}^z,m_{zz})^T,\\
m^{\epsilon}&=&(m_{x0},m_{y0},m_{0x},m_{0y},m_{xz},m_{zx},m_{yz},m_{zy})^T,\nonumber
\end{eqnarray}
where
$$
m_{14}^{\alpha}={\rm
tr}(\Omega_{14}^{\alpha}\rho),\,m_{23}^{\beta}={\rm
tr}(\Omega_{23}^{\beta}\rho),\,\,\alpha,\beta=x,y,z,$$
$$m_{ij}={\rm tr}\left[\left(\frac{1}{2}\sigma_i^{(1)}\sigma_j^{(2)}\right)\rho\right],\,\,i,j=0,x,y,z,
$$
and $\sigma_0^{(i)}=I_{2 \times 2},\,i=1,2$ are the $2\times 2$
identity operators acting on the $i$-th atom. Then,
(\ref{Coherence vector equation of independent amplitude damping
decoherence channel}) can be grouped into:
\begin{eqnarray}\label{Variable equation of independent amplitude damping decoherence channel}
\dot{m}^p&=&\sum_{i=1}^4 u_i O_i^{\eta} m^{\eta}+D^p m^p,\nonumber\\
\dot{m}^{\eta}&=&\sum_{i=1}^4u_i(-O_i^{\eta\,T})m^p+D^{\eta}m^{\eta}+g^{\eta},\\
\dot{m}^{\epsilon}&=&\sum_{i=1}^4 u_i O_i^{\epsilon}
m^{\epsilon}+D^{\epsilon}m^{\epsilon},\nonumber
\end{eqnarray}
where \setlength{\mathindent}{2.5cm}
\begin{eqnarray}\label{Matrices in the coherence vector picture}
&u_1=8\mu_1\cos\phi_1,\,u_2=8\mu_1\sin\phi_1,\,u_3=8\mu_2\cos\phi_2,\,u_4=-8\mu_2\sin\phi_2,&\nonumber\\
&D^p=-4\Gamma\left(%
\begin{array}{cccc}
  1 &  &  &  \\
   & 1 &  &  \\
   &  & 1 &  \\
   &  &  & 1 \\
\end{array}%
\right),\,\,O_1^{\eta}=\left(%
\begin{array}{ccc}
  0 & 0 & 0 \\
  -1 & 0 & 0 \\
  0 & 0 & 0 \\
  0 & 0 & 0 \\
\end{array}%
\right),\nonumber\\
&O_2^{\eta}=\left(%
\begin{array}{ccc}
  1 & 0 & 0 \\
  0 & 0 & 0 \\
  0 & 0 & 0 \\
  0 & 0 & 0 \\
\end{array}%
\right),\,\,O_3^{\eta}=\left(%
\begin{array}{ccc}
  0 & 0 & 0 \\
  0 & 0 & 0 \\
  0 & 0 & 0 \\
  0 & -1 & 0 \\
\end{array}%
\right),\,\,O_4^{\eta}=\left(%
\begin{array}{ccc}
  0 & 0 & 0 \\
  0 & 0 & 0 \\
  0 & 1 & 0 \\
  0 & 0 & 0 \\
\end{array}%
\right),\nonumber\\
&D^{\eta}=-4\Gamma\left(%
\begin{array}{ccc}
  1 & 0 & 0 \\
  0 & 1 & 0 \\
  -\sqrt{2} & 0 & 2 \\
\end{array}%
\right),\,\,g^{\eta}=\left(%
\begin{array}{c}
  2\sqrt{2}\Gamma \\
  0 \\
  0 \\
\end{array}%
\right).&
\end{eqnarray}
$O_i^{\epsilon}$ are all skew-symmetric matrices and
$D^{\epsilon}<0$, thereby
$$
\frac{d}{dt}((m^{\epsilon})^Tm^{\epsilon})=(m^{\epsilon})^TD^{\epsilon}m^{\epsilon}<0,\quad\forall
m^\epsilon\neq 0,
$$
which implies that $m^{\epsilon}\rightarrow 0$ when
$t\rightarrow\infty$.

With simple calculations, the following stationary solution can be
obtained for Eq. (\ref{Variable equation of independent amplitude
damping decoherence channel}):
$$m^{\epsilon}=0,\,m_{23}^x(\infty)=m_{23}^y(\infty)=m_{23}^z(\infty)=0,$$
$$m_{14}^x(\infty)=\frac{\sqrt{2}\mu_1\Gamma}{4\mu_1^2+\Gamma^2}\cos\left(\phi_1-\frac{\pi}{2}\right),$$
$$m_{14}^y(\infty)=\frac{\sqrt{2}\mu_1\Gamma}{4\mu_1^2+\Gamma^2}\sin\left(\phi_1-\frac{\pi}{2}\right),$$
$$m_{14}^z(\infty)=\sqrt{2}m_{zz}(\infty)=\frac{\sqrt{2}\Gamma^2}{8\mu_1^2+2\Gamma^2},$$
from which we can obtain the corresponding decomposition
(\ref{Controlled stationary state under independent amplitude
damping decoherence channels}) of the stationary state
$\rho_{\infty}$.

Going back to the density matrix, one can find that the stationary
state has the following form: \setlength{\mathindent}{5cm}
\begin{equation}\label{Special form of the stationary state}
\rho_{\infty}=\left(%
\begin{array}{cccc}
  a &  &  & w \\
   & b & z &  \\
   & z^* & c &  \\
  w^* &  &  & d \\
\end{array}%
\right),
\end{equation}
whose concurrence can be analytically solved to
be~\cite{Yu1,Yu2,Yu3}: \setlength{\mathindent}{4cm}
\begin{equation}\label{Concurrence of special stationary state}
C(\rho_{\infty})=2\max\{|w|-\sqrt{bc},|z|-\sqrt{ad},0\}.
\end{equation}
The above equation leads to the stationary concurrence
$C(\rho_{\infty})$ given in Eq. (\ref{Stationary concurrence for
independent amplitude damping decoherence channel}).

\section{Proof of the Equations (\ref{Stationary state of the uncontrolled collective amplitude damping decoherence channel})--(\ref{Controlled stationary concurrence under collective amplitude damping decoherence})}\label{Proof of the results for the
collective amplitude damping decoherence}

Similar to what we have done in \ref{Proof of the results for the
independent amplitude damping decoherence}, the controlled master
equation (\ref{Master equation of collective amplitude damping
decoherence channel}) can be grouped into the following control
equations in the coherence vector picture:
\setlength{\mathindent}{3cm}
\begin{eqnarray}\label{Coherent vector equation for collective amplitude damping decoherence}
\dot{m}_{14}^x&=&8\mu_1\sin\theta_1 m_{14}^z -4\Gamma m_{14}^x,\nonumber\\
\dot{m}_{14}^y&=&-8\mu_1\cos\theta_1 m_{14}^z-4\Gamma m_{14}^y,\nonumber\\
\dot{m}_{14}^z&=&8\mu_1\cos\theta_1 m_{14}^y-8\mu_1\sin\theta_1
m_{14}^x-4\Gamma
m_{14}^z+2\sqrt{2}\Gamma+4\Gamma m_{23}^x,\nonumber\\
\dot{m}_{23}^x&=&-4\Gamma m_{23}^x-4\Gamma
m_{14}^z+4\sqrt{2}\Gamma m_{zz},\nonumber\\
\dot{m}_{23}^y&=&-8(\mu_2+\eta_0) m_{23}^z-4\Gamma m_{23}^y,\nonumber\\
\dot{m}_{23}^z&=&8(\mu_2+\eta_0) m_{23}^y-4\Gamma
m_{23}^z,\nonumber\\
\dot{m}_{zz}&=&4\sqrt{2}\Gamma m_{14}^z-8\Gamma
m_{zz}+4\sqrt{2}\Gamma m_{23}^x.
\end{eqnarray}
As well as the independent amplitude damping decoherence model,
the sub-vector $m^{\epsilon}=(m_{x0},\cdots,m_{zy})^T$ always goes
to zero when $t\rightarrow\infty$, and it will not affect
$C(\rho_{\infty})$, so we will not discuss $m^{\epsilon}(t)$ here.
The fourth and the last equations in Eq.~(\ref{Coherent vector
equation for collective amplitude damping decoherence}) implies a
conservation law:
$$
m_{zz}(t)+\sqrt{2}m_{23}^x(t)\equiv
m_{zz}(t_0)+\sqrt{2}m_{23}^x(t_0)\triangleq2\kappa-\frac{3}{2}.
$$
Substituting $\mu_1=\mu_2=0$ into Eq.~(\ref{Coherent vector
equation for collective amplitude damping decoherence}), we have
the following uncontrolled stationary solution:
$$m_{14}^x(\infty)=m_{14}^y(\infty)=m_{23}^y(\infty)=m_{23}^z(\infty)=0,$$
$$m_{14}^z(\infty)=\frac{\sqrt{2}}{2}\kappa,\,\,\,m_{23}^x(\infty)=\frac{\sqrt{2}}{2}(\kappa-1),\,\,\,m_{zz}(\infty)=\kappa-\frac{1}{2},$$
from which the decomposition (\ref{Stationary state of the
uncontrolled collective amplitude damping decoherence channel})
can be obtained and $C(\rho^u_{\infty})=1-\kappa$.

Further, we can obtain the controlled stationary solution of
Eq.~(\ref{Coherent vector equation for collective amplitude
damping decoherence}):
$$m_{14}^x(\infty)=\frac{\sqrt{2}\mu_1\Gamma\kappa}{\Gamma^2+3\mu_1^2}\cos\left(\phi_1-\frac{\pi}{2}\right),\,\,\,m_{14}^y(\infty)=\frac{\sqrt{2}\mu_1\Gamma\kappa}{\Gamma^2+3\mu_1^2}\sin\left(\phi_1-\frac{\pi}{2}\right),$$
$$m_{23}^y(\infty)=m_{23}^z(\infty)=0,\,\,\,m_{14}^z(\infty)=\frac{\sqrt{2}\kappa\Gamma^2}{2(\Gamma^2+3\mu_1^2)},$$
$$m_{zz}(\infty)=\kappa-\frac{1}{2}-\frac{\mu_1^2\kappa}{\Gamma^2+3\mu_1^2},\,\,\,m_{23}^x(\infty)=\frac{\sqrt{2}}{2}\left(\kappa-1+\frac{\mu_1^2\kappa}{\Gamma^2+3\mu_1^2}\right),$$
which leads to the decomposition (\ref{Stationary state of the
controlled collective amplitude damping decoherence channel}).
Further, it can be calculated that
$$
C(\rho_{\infty})=\max\left\{F_1(\mu_1),F_2(\mu_2),0\right\},
$$
where
$$
F_1(\mu_1)=-\kappa+1+\frac{\mu_1^2}{\Gamma^2+3\mu_1^2}\kappa-\frac{2\mu_1\sqrt{\Gamma^2+\mu_1^2}}{\Gamma^2+3\mu_1^2}\kappa,
$$
$$
F_2(\mu_1)=\frac{(\Gamma^2+2\mu_1^2+2\Gamma\mu_1)}{\Gamma^2+3\mu_1^2}\kappa-1.
$$ It is easy to verify that $F_1(\mu_1)$ monotonically decreases
when the control parameter $\mu_1$ increases, thereby
$$
F_1(\mu_1)\leq F_1(0)=1-\kappa=C(\rho^u_{\infty}).
$$
Also, we can obtain that
$$
F_2(\mu_1)\geq 1-\kappa=C(\rho^u_{\infty})\geq 0,
$$
when $\mu_1$ satisfies Eq. (\ref{Interval of the control
parameter}). In conclusion, we arrive at
$$
C(\rho_{\infty})=F_2(\mu_1)\geq 1-\kappa=C(\rho^u_{\infty}).
$$
\\[0.2cm]
\noindent\textbf{\large Acknowledgments}
\\[0.1cm]

The authors would like to thank Dr. Yu-xi Liu for helpful
discussions and valuable advice. This research was supported in
part by the National Natural Science Foundation of China under
Grant Nos. 60704017, 60433050, 60635040, 60674039 and China
Postdoctoral Science Foundation. T. J. Tarn would also like to
acknowledge partial support from the U.S. Army Research Office
under Grant W911NF-04-1-0386.
\\[0.1cm]

\end{document}